\documentclass[aps,preprint,nofootinbib,superscriptaddress]{revtex4}%
\usepackage{amssymb}
\usepackage{amsmath}
\usepackage{epsfig}
\usepackage{amsfonts}
\usepackage{graphicx}
\usepackage{hyperref}%
\providecommand{\U}[1]{\protect\rule{.1in}{.1in}}
\begin{document}
\title{The String BCJ Relations Revisited and Extended Recurrence relations of
Nonrelativistic String Scattering Amplitudes}
\author{Sheng-Hong Lai}
\email{xgcj944137@gmail.com}
\affiliation{Department of Electrophysics, National Chiao-Tung University, Hsinchu, Taiwan, R.O.C.}
\author{Jen-Chi Lee}
\email{jcclee@cc.nctu.edu.tw}
\affiliation{Department of Electrophysics, National Chiao-Tung University, Hsinchu, Taiwan, R.O.C.}
\author{Yi Yang}
\email{ihtsai@math.ntu.edu.tw}
\affiliation{Department of Electrophysics, National Chiao-Tung University, Hsinchu, Taiwan, R.O.C.}
\author{}
\date{\today }

\begin{abstract}
We review and extend high energy four point string BCJ relations in both the
fixed angle and Regge regimes. We then give an explicit proof of four point
string BCJ relations for all energy. This calculation provides an alternative
proof of the one based on monodromy of integration in string amplitude
calculation. In addition, we calculate both $s-t$ and $t-u$ channel
nonrelativistic low energy string scattering amplitudes of three tachyons and
one higher spin string state at arbitrary mass levels. We discover that the
mass and spin dependent nonrelativistic string BCJ relations can be expressed
in terms of Gauss hypergeometry functions. As an application, for each fixed
mass level $N,$ we derive extended recurrence relations among nonrelativistic
low energy string scattering amplitudes of string states with different spins
and different channels.

\end{abstract}
\maketitle
\tableofcontents
%

%TCIMACRO{\TeXButton{equation number}{\setcounter{equation}{0}
%\renewcommand{\theequation}{\arabic{section}.\arabic{equation}}}}%
%BeginExpansion
\setcounter{equation}{0}
\renewcommand{\theequation}{\arabic{section}.\arabic{equation}}%
%EndExpansion

\section{\bigskip Introduction}

Inspired by Witten's seminal paper published in 2004 \cite{witten}, there have
been tremendous developments on calculations of higher point and higher loop
Yang-Mills and gravity field theory amplitudes \cite{JA}. Many new ideas and
techniques have been proposed and suggested on this interesting and important
subject. On the other hand, string theory amplitudes have been believed to be
closely related to these new results derived in field theory amplitudes. One
interesting example was the gauge field theory BCJ relations for
color-stripped amplitudes proposed in 2008 \cite{BCJ} and their string origin
or the string BCJ relations suggested in 2009 \cite{stringBCJ,stringBCJ2}.
These field theory BCJ relations can be used to reduce the number of
independent $n$-point color-ordered gauge field theory amplitudes from
$(n-2)!,$ as suggested by the KK relations \cite{KK,KK2}, to $(n-3)!$.

On the other hand, a less known historically independent development of the
"string BCJ relations" was from the string theory side \textit{without}
refering to the field theory BCJ relations. This was the discovery of the four
point string BCJ relations in the high energy fixed angle or hard string
scattering (HSS) limit in 2006 \cite{Closed}. Moreover, one can combine these
string BCJ relations with the infinite linear relations among HSS amplitudes
conjectured by Gross \cite{GM,Gross,GrossManes} and discovered in
\cite{ChanLee1,ChanLee2,CHL,PRL, CHLTY,susy} to form the \textit{extended
linear relations }in the HSS limit. For a recent review, see \cite{review}. In
constrast to the field theory BCJ relation, these extended linear relations
relate HSS amplitudes of string states with different spins and different
channels, and can be used to reduce the number of independent HSS amplitudes
from $\infty$ down to $1$.

Historically, the motivation to probe string BCJ relations in this context was
the calculation of \textit{closed} HSS amplitudes \cite{Closed} by using the
KLT relations \cite{KLT}. Indeed, it was found \cite{Closed} that the saddle
point calculation of open HSS amplitudes was applicable only for the $t-u$
channel, but not reliable for the $s-t$ channel, neither for the closed HSS
amplitude calculation. In addition, it was also pointed out
\cite{Closed,Closed2} that the prefactor $\frac{\sin\left(  \pi u/2\right)
}{\sin\left(  \pi s/2\right)  }$ in the string BCJ relations, which was
missing in the literature \cite{zeros,GM,Gross,GrossManes} for the HSS
amplitude calculations, had important physical interpretations. The poles give
infinite number of resonances in the string spectrum and zeros give the
coherence of string scatterings. These poles and zeros survive in the HSS
limit and can not be ignored. Presumably, the prefactor triggers the failure
of saddle point calculations mentioned above.

To calculate the closed HSS by KLT relation, one needed to calculate both
$s-t$ and $t-u$ channel HSS amplitudes. In constrast to the saddle point
method used in the $t-u$ channel, for the $s-t$ channel HSS amplitudes at each
fixed mass level $N,$ one first used a direct method to calculate the HSS
amplitude of the leading trajectory string state, and then extended the result
to other string states by using high energy symmetry of string theory or
infinite linear relations among HSS amplitudes of different string states at
each mass level $N$. As a result, the string BCJ relations in the HSS limit
can be derived \cite{Closed}. All these HSS amplitude calculations for $s-t$
and\ $t-u$ channels and the related string BCJ relations in the HSS limit were
inspired by Gross famous conjectures on high energy symmetries of string
theory \cite{Gross}, and thus were independent and different from the field
theory BCJ motivation discussed above.

In this paper, we will follow up the development on the string theory side of
string amplitude calculations. In section II, we will first review the
extended linear relations\textit{ }in the HSS limit discussed in
\cite{Closed}. We will also work out the corresponding \textit{extended
recurrence relations }\cite{LY,LY2014} of the Regge string scattering (RSS)
amplitudes \cite{KLY}. Similar to the extended linear relations\textit{ }in
the HSS limit discussed above, the extended recurrence relations can be used
to reduce the number of independent RSS amplitudes from $\infty$ down to $1$.

We will then give an explicit proof of the string BCJ relations in section III
by directly calculating $s-t$ and\ $t-u$ channel string scattering amplitudes
for arbitrary four string states. In constrast to the proof based on monodromy
of integration with constraints on the kinematic regime given in
\cite{stringBCJ} without calculating string amplitudes, our explicit string
amplitude calculation puts no constraints on the kinematic regime. In section
IV, we will calculate the level dependent and the extended recurrence
relations of low energy string scattering amplitudes\textit{ }in the
nonrelativistic limit. The existence of recurrence relations for low energy
nonrelativistic string scattering (NSS) amplitudes come as a surprise from
Gross point of view on HSS limit.

In the calculations of low energy extended recurrence relations in section IV,
we will take the NSS limit or $|\vec{k_{2}}|<<M_{S}$ limit to calculate the
mass level and spin dependent low energy NSS amplitudes. In constrast to the
zero slope $\alpha^{\prime}$ limit used in the literature to calculate the
massless Yang-Mills couplings \cite{ymzero1,ymzero2} for superstring and the
three point $\varphi^{3}$ scalar field coupling \cite{Bzero1,Bzero2,Bzero3}
for the bosonic string, we found it appropriate to take the nonrelativistic
limit in calculating low energy string scattering amplitudes for string states
with both higher spins and finite mass gaps. A brief conclusion will be given
in section V. In the following in section II, we\ first review historically
two independnt developments of the string BCJ relations from field theory and
from string theory point of views.%

%TCIMACRO{\TeXButton{equation number}{\setcounter{equation}{0}
%\renewcommand{\theequation}{\arabic{section}.\arabic{equation}}}}%
%BeginExpansion
\setcounter{equation}{0}
\renewcommand{\theequation}{\arabic{section}.\arabic{equation}}%
%EndExpansion

\section{Review of High energy String BCJ}

The four point BCJ relations \cite{BCJ} for Yang-Mills gluon color-stripped
scattering amplitudes $A$ were pointed out and calculated in 2008 to be
\begin{subequations}
\label{BCJ}%
\begin{align}
tA(k_{1},k_{4},k_{2},k_{3})-sA(k_{1},k_{3},k_{4},k_{2})  &  =0,\\
sA(k_{1},k_{2},k_{3},k_{4})-uA(k_{1},k_{4},k_{2},k_{3})  &  =0,\\
\text{ }uA(k_{1},k_{3},k_{4},k_{2})-tA(k_{1},k_{2},k_{3},k_{4})  &  =0,
\end{align}
which relates field theory scattering amplitudes in the $s$, $t$ and $u$
channels. In the rest of this paper, we will discuss the relation for $s$ and
$u$ channel amplitudes only. Other relations can be similarly addressed.

\subsection{ Hard String Scatterings}

For string theory, in constrast to the field theory BCJ relations, one has to
deal with scattering amplitudes of infinite number of higher spin string
states. The first "string BCJ relation" discovered was the four point string
BCJ relation in the HSS limit \cite{Closed} worked out in 2006. For the
tachyon state, one can express the open string $s-t$ channel amplitude in
terms of the $t-u$ channel amplitude \cite{Closed}
\end{subequations}
\begin{align}
T_{\text{open}}^{\left(  4\text{-tachyon}\right)  }\left(  s,t\right)   &
=\frac{\Gamma\left(  -\frac{s}{2}-1\right)  \Gamma\left(  -\frac{t}%
{2}-1\right)  }{\Gamma\left(  \frac{u}{2}+2\right)  }\nonumber\\
&  =\frac{\sin\left(  \pi u/2\right)  }{\sin\left(  \pi s/2\right)  }%
\frac{\Gamma\left(  -\frac{t}{2}-1\right)  \Gamma\left(  -\frac{u}%
{2}-1\right)  }{\Gamma\left(  \frac{s}{2}+2\right)  }\nonumber\\
&  \equiv\frac{\sin\left(  \pi u/2\right)  }{\sin\left(  \pi s/2\right)
}T_{\text{open}}^{\left(  4\text{-tachyon}\right)  }\left(  t,u\right)
\label{teq}%
\end{align}
where\ we have used the well known formula
\begin{equation}
\Gamma\left(  x\right)  =\frac{\pi}{\sin\left(  \pi x\right)  \Gamma\left(
1-x\right)  }. \label{math}%
\end{equation}
The\ string BCJ relation for tachyon derived above is valid for all energies.

For the $N$-point open string \textit{tachyon} amplitudes $B_{N}$ of
Koba-Nielson, some authors in the early days of dual models, see for example
\cite{Plahte}, discussed symmetry relations among $B_{N}$ functions with
different cyclic order external momenta by using monodromy of countour
integration of the amplitudes. However, no discussion was addressed for string
amplitudes with infinite number of \textit{higher spin} massive string states,
on which we will discuss next.

Since there is no reliable saddle point to calculate $s-t$ channel HSS
amplitudes, for $all$ other higher spin string states at arbitrary mass
levels, one first calculates the $s-t$ channel scattering amplitude with
$V_{2}=\alpha_{-1}^{\mu_{1}}\alpha_{-1}^{\mu_{2}}..\alpha_{-1}^{\mu_{n}}%
\mid0,k>$, the highest spin state at mass level $M_{2}^{2}$ $=2(N-1),$ and
three tachyons $V_{1,3,4}$ as \cite{CHLTY,Closed}%
\begin{equation}
\mathcal{T}_{N;st}^{\mu_{1}\mu_{2}\cdot\cdot\mu_{n}}=\sum_{l=0}^{N}%
(-)^{l}\binom{N}{l}B\left(  -\frac{s}{2}-1+l,-\frac{t}{2}-1+N-l\right)
k_{1}^{(\mu_{1}}..k_{1}^{\mu_{n-l}}k_{3}^{\mu_{n-l+1}}..k_{3}^{\mu_{N})}.
\label{B}%
\end{equation}
The\ corresponding $t-u$ channel open string scattering amplitude can be
calculated to be \cite{Closed}%
\begin{equation}
\mathcal{T}_{N;tu}^{\mu_{1}\mu_{2}\cdot\cdot\mu_{n}}=\sum_{l=0}^{N}\binom
{N}{l}B\left(  -\frac{t}{2}+N-l-1,-\frac{u}{2}-1\right)  k_{1}^{(\mu_{1}%
}..k_{1}^{\mu_{N-l}}k_{3}^{\mu_{N-l+1}}k_{3}^{\mu_{N})}.
\end{equation}
The HSS limit of the string BCJ relation for these amplitudes was worked out
to be \cite{Closed}%
\begin{equation}
\mathcal{T}_{N}(s,t)\simeq(-)^{N}\frac{\sin\left(  \pi u/2\right)  }%
{\sin\left(  \pi s/2\right)  }\mathcal{T}_{N}(t,u) \label{HBCJ}%
\end{equation}
where
\begin{align}
\mathcal{T}_{N}(t,u)  &  \simeq\sqrt{\pi}(-1)^{N-1}2^{-N}E^{-1-2N}\left(
\sin\frac{\phi}{2}\right)  ^{-3}\left(  \cos\frac{\phi}{2}\right)
^{5-2N}\nonumber\\
&  \cdot\exp\left[  -\frac{t\ln t+u\ln u-(t+u)\ln(t+u)}{2}\right]  \label{tu}%
\end{align}

Note that unlike the case of tachyon, this relation was proved only for HSS
limit. The next key step was that the result of Eq.(\ref{HBCJ}) can be
generalized to the case of three tachyons and one arbitrary string states
\cite{PRL,CHLTY}, and then to the case of four arbitrary string states. This
generalization was based on the important result that, at each fixed mass
level $N,$ the high energy fixed angle string scattering amplitudes for states
differ from leading Regge trajectory higher spin state in the second vertex
are all proportional to each other \cite{PRL,CHLTY}
\begin{equation}
\frac{T_{st}^{(N,2m,q)}}{T_{st}^{(N,0,0)}}\simeq\frac{T_{tu}^{(N,2m,q)}%
}{T_{tu}^{(N,0,0)}}=\left(  -\frac{1}{M}\right)  ^{2m+q}\left(  \frac{1}%
{2}\right)  ^{m+q}(2m-1)!!. \label{HS}%
\end{equation}
Here $\mathcal{T}_{N}(t,u)=T_{tu}^{(N,0,0)}$ for the case of three tachyons
and one tensor, and $T^{(N,2m,q)}$ represents leading order hard open string
scattering amplitudes with three arbitrary string states and one higher spin
string state of the following form \cite{PRL,CHLTY}
\begin{equation}
\left\vert N,2m,q\right\rangle \equiv(\alpha_{-1}^{T})^{N-2m-2q}(\alpha
_{-1}^{L})^{2m}(\alpha_{-2}^{L})^{q}|0,k\rangle\label{GR}%
\end{equation}
where the polarizations of the higher spin string state with momentum $k_{2}$
on the scattering plane were defined to be $e^{P}=\frac{1}{M_{2}}%
(E_{2},\mathrm{k}_{2},0)=\frac{k_{2}}{M_{2}}$, $e^{L}=\frac{1}{M_{2}%
}(\mathrm{k}_{2},E_{2},0)$ and $e^{T}=(0,0,1)$, and we have omitted possible
tensor indices of the other three string states. This high energy symmetry of
string theory was first conjectured by Gross in 1988 \cite{Gross} and was
explicitly proved in \cite{ChanLee1,ChanLee2,PRL, CHLTY,susy}.

Finally, the string BCJ relations for arbitrary string states in the hard
scattering limit can be written as \cite{Closed}%
\begin{equation}
T_{st}^{(N,2m,q)}\simeq(-)^{N}\frac{\sin\left(  \pi u/2\right)  }{\sin\left(
\pi s/2\right)  }T_{tu}^{(N,2m,q)}=\frac{\sin\left(  \pi k_{2}.k_{4}\right)
}{\sin\left(  \pi k_{1}.k_{2}\right)  }T_{tu}^{(N,2m,q)}. \label{HBCJ2}%
\end{equation}
Eq.(\ref{HS}) and Eq.(\ref{HBCJ2}) can be combined to form the
\textit{extended linear relations in the HSS limit}. These relations relate
string scattering amplitudes of string states with different spins and
different channels in the HSS limit, and can be used to reduce the infinite
number of independent hard string scattering amplitudes from $\infty$ down to
$1$.

Note that historically the motivation to probe string theory BCJ relations in
this context was the calculation of high energy closed string scattering
amplitudes \cite{Closed} by using the KLT relations \cite{KLT}. Indeed, it was
found that the saddle point calculations of high energy fixed angle open
string scattering amplitudes were available only for the $t-u$ channel, but
not reliable for the $s-t$ channel neither for the closed string high energy
amplitude calculation \cite{Closed}. So it was important to use other method
to express $s-t$ channel hard string scattering amplitudes in terms of $t-u$
channel HSS amplitudes.

The factor in Eq.(\ref{HBCJ2}) $\frac{\sin\left(  \pi u/2\right)  }%
{\sin\left(  \pi s/2\right)  }$ which was missing in the literature
\cite{GM,zeros} has important physical interpretations \cite{Closed}. The
presence of poles give infinite number of resonances in the string spectrum
and zeros give the coherence of string scatterings. These poles and zeros
survive in the high energy limit and can not be dropped out. Presumably, the
factor in the string BCJ relation in Eq.(\ref{HBCJ2}) triggers the failure of
saddle point calculation in the $s-t$ channel.

The two relations in Eq.(\ref{teq}) and Eq.(\ref{HBCJ2}) can be written as the
four point \textit{string BCJ relation} which are valid to all energies as%
\begin{equation}
A_{st}=\frac{\sin\left(  \pi k_{2}.k_{4}\right)  }{\sin\left(  \pi k_{1}%
k_{2}\right)  }A_{tu} \label{BCJ3}%
\end{equation}
\textit{if} one can generalize the proof of Eq.(\ref{HBCJ2}) to all energies.
This was done in a paper based on monodromy of integration in string amplitude
calculation published in 2009 \cite{stringBCJ}. However, the explicit forms of
the amplitudes $A_{st}$ and $A_{tu}$ were \textit{not} calculated in
\cite{stringBCJ} and the extended linear relations were not addressed there.
In the next section, we will provide an alternative proof of this string BCJ relation.

\subsection{Regge string scatterings}

Another interesting regime of the string BCJ relation in Eq.(\ref{BCJ3}) was
the Regge regime. The $s-t$ channel RSS amplitude of three tachyons and one
higher spin state in Eq.(\ref{GR}) was calculated to be \cite{KLY}%
\begin{align}
R_{st}^{(n,2m,q)}=  &  B\left(  -1-\frac{s}{2},-1-\frac{t}{2}\right)
\sqrt{-t}^{n-2m-2q}\left(  \frac{1}{2M_{2}}\right)  ^{2m+q}\nonumber\\
\cdot &  2^{2m}(\tilde{t}^{\prime})^{q}U\left(  -2m\,,\,\frac{t}%
{2}+2-2m\,,\,\frac{\tilde{t}^{\prime}}{2}\right)  \label{13}%
\end{align}
where $\tilde{t}^{\prime}=t+M_{2}^{2}-M_{3}^{2}$ and $U$ is the Kummer
function of the second kind. The corrsponding $t-u$ channel amplitude was
calculated to be \cite{Closed2}%
\begin{align}
R_{tu}^{(n,2m,q)}  &  =(-)^{n}B(-1-\frac{t}{2},-1-\frac{u}{2})(\sqrt{-{t}%
})^{n-2m-2q}\left(  \frac{1}{2M_{2}}\right)  ^{2m+q}\nonumber\\
&  \cdot2^{2m}(\tilde{t}^{\prime})^{q}U\left(  -2m\,,\,\frac{t}{2}%
+2-2m\,,\,\frac{\tilde{t}^{\prime}}{2}\right)  . \label{18}%
\end{align}
We can calculate the ratio of the two amplitudes as following%
\begin{align}
\frac{R_{st}^{(n,2m,q)}}{R_{tu}^{(n,2m,q)}}  &  =\left(  -1\right)  ^{n}%
\frac{B\left(  -\frac{s}{2}-1,-\frac{t}{2}-1\right)  }{B\left(  -\frac{t}%
{2}-1,-\frac{u}{2}-1\right)  }\nonumber\\
&  =\frac{\left(  -1\right)  ^{n}\Gamma\left(  -\frac{s}{2}-1\right)
\Gamma\left(  \frac{s}{2}+2-n\right)  }{\Gamma\left(  -\frac{u}{2}-1\right)
\Gamma\left(  \frac{u}{2}+2-n\right)  }\nonumber\\
&  =\frac{\left(  -1\right)  ^{n}\Gamma\left(  -\frac{s}{2}-1\right)
\Gamma\left(  \frac{s}{2}+2\right)  }{\Gamma\left(  -\frac{u}{2}-1\right)
\Gamma\left(  \frac{u}{2}+2\right)  }\cdot\frac{\left(  \frac{u}%
{2}+2-n\right)  \cdots\left(  \frac{u}{2}+1\right)  }{\left(  \frac{s}%
{2}+2-n\right)  \cdots\left(  \frac{s}{2}+1\right)  }.
\end{align}
One can now take the Regge limit, $t=$ fixed and $s\sim-u\rightarrow\infty$ to
get%
\begin{equation}
\frac{R_{st}^{(n,2m,q)}}{R_{tu}^{(n,2m,q)}}\simeq\frac{\left(  -1\right)
^{n}\Gamma\left(  -\frac{s}{2}-1\right)  \Gamma\left(  \frac{s}{2}+2\right)
}{\Gamma\left(  -\frac{u}{2}-1\right)  \Gamma\left(  \frac{u}{2}+2\right)
}\cdot\left(  -1\right)  ^{n}=\frac{\sin\pi(\frac{u}{2}+2)}{\sin\pi(\frac
{s}{2}+2)}=\frac{\sin\pi\left(  k_{2}\cdot k_{4}\right)  }{\sin\pi\left(
k_{1}\cdot k_{2}\right)  },
\end{equation}
which can be written as%
\begin{equation}
R_{st}^{(n,2m,q)}\simeq\frac{\sin\left(  \pi k_{2}.k_{4}\right)  }{\sin\left(
\pi k_{1}.k_{2}\right)  }R^{(n,2m,q)}(t,u), \label{RR1}%
\end{equation}
and is consistent with the string BCJ relation in Eq.(\ref{BCJ3}).

The result in Eq.(\ref{RR1}) can be generalized to the general leading order
Regge string states at each fixed mass level $N=\sum_{n,m,l>0}np_{n}%
+mq_{m}+lr_{l}$ \cite{KLY}%
\begin{equation}
\left\vert p_{n},q_{m},r_{l}\right\rangle =\prod_{n>0}(\alpha_{-n}^{T}%
)^{p_{n}}\prod_{m>0}(\alpha_{-m}^{P})^{q_{m}}\prod_{l>0}(\alpha_{-l}%
^{L})^{r_{l}}|0,k\rangle. \label{RR}%
\end{equation}
For this general case, the $s-t$ channel of the scattering amplitude in the
Regge limit was calculated to be \cite{LY2014}%

\begin{align}
R_{st}^{\left\vert p_{n},q_{m},r_{l}\right\rangle }=  &  \prod_{n=1}\left[
\left(  n-1\right)  !\sqrt{-t}\right]  ^{p_{n}}\prod_{m=1}\left[  \left(
m-1\right)  !\frac{\tilde{t}}{2M_{2}}\right]  ^{q_{m}}\prod_{l=1}\left[
\left(  l-1\right)  !\frac{\tilde{t}^{\prime}}{2M_{2}}\right]  ^{r_{l}%
}\nonumber\\
\cdot &  F_{1}\left(  -\frac{t}{2}-1;-q_{1},-r_{1};\frac{s}{2};-\frac
{s}{\tilde{t}},-\frac{s}{\tilde{t}^{\prime}}\right)  B\left(  -\frac{s}%
{2}-1,-\frac{t}{2}-1\right)
\end{align}
where $F_{1}$ is the first Appell function. On the other hand, one can
calculate the $t-u$ channel amplitude, which was missing in \cite{LY2014}, as following%

\begin{align}
R_{tu}^{\left\vert p_{n},q_{m},r_{l}\right\rangle }=  &  \int_{1}^{\infty
}dx\cdot x^{k_{1}\cdot k_{2}}(x-1)^{k_{2}\cdot k_{3}}\left[  \frac{k_{1}^{P}%
}{x}-\frac{k_{3}^{P}}{1-x}\right]  ^{q_{1}}\left[  \frac{k_{1}^{L}}{x}%
-\frac{k_{3}^{L}}{1-x}\right]  ^{r_{l}}\nonumber\\
\cdot &  \prod_{n=1}\left[  -\frac{\left(  n-1\right)  !k_{3}^{T}}{\left(
1-x\right)  ^{n}}\right]  \prod_{m=2}\left[  -\frac{\left(  m-1\right)
!k_{3}^{P}}{\left(  1-x\right)  ^{m}}\right]  \prod_{l=2}\left[
-\frac{\left(  l-1\right)  !k_{3}^{L}}{\left(  1-x\right)  ^{l}}\right]
\nonumber\\
=  &  \prod_{n=1}\left[  \left(  n-1\right)  !\sqrt{-t}\right]  ^{p_{n}}%
\prod_{m=1}\left[  \left(  m-1\right)  !\frac{\tilde{t}}{2M_{2}}\right]
^{q_{m}}\prod_{l=1}\left[  \left(  l-1\right)  !\frac{\tilde{t}^{\prime}%
}{2M_{2}}\right]  ^{r_{l}}\nonumber\\
\cdot &  \sum_{i}^{q_{1}}\binom{q_{1}}{i}\left(  -\frac{s}{\tilde{t}}\right)
^{i}\sum_{j}^{r_{1}}\binom{r_{1}}{j}\left(  -\frac{s}{\tilde{t}^{\prime}%
}\right)  ^{j}\left(  -1\right)  ^{k_{2}\cdot k_{3}}\nonumber\\
\cdot &  \int_{1}^{\infty}dx\cdot x^{-\frac{s}{2}-2+N-i-j}(1-x)^{-\frac{t}%
{2}-2+i+j}.
\end{align}
We can do the change of variable $y=\frac{x-1}{x}$ to get%

\begin{align}
R_{tu}^{\left\vert p_{n},q_{m},r_{l}\right\rangle }=  &  \left(  -1\right)
^{N}\prod_{n=1}\left[  \left(  n-1\right)  !\sqrt{-t}\right]  ^{p_{n}}%
\prod_{m=1}\left[  \left(  m-1\right)  !\frac{\tilde{t}}{2M_{2}}\right]
^{q_{m}}\prod_{l=1}\left[  \left(  l-1\right)  !\frac{\tilde{t}^{\prime}%
}{2M_{2}}\right]  ^{r_{l}}\nonumber\\
\cdot &  \sum_{i}^{q_{1}}\binom{q_{1}}{i}\left(  -\frac{s}{\tilde{t}}\right)
^{i}\sum_{j}^{r_{1}}\binom{r_{1}}{j}\left(  -\frac{s}{\tilde{t}^{\prime}%
}\right)  ^{j}\left(  -1\right)  ^{-i-j}\int_{0}^{1}dy\cdot y^{\frac{-t}%
{2}-2+i+j}\left(  1-y\right)  ^{\frac{-u}{2}-2}\nonumber\\
=  &  \left(  -1\right)  ^{N}\prod_{n=1}\left[  \left(  n-1\right)  !\sqrt
{-t}\right]  ^{p_{n}}\prod_{m=1}\left[  \left(  m-1\right)  !\frac{\tilde{t}%
}{2M_{2}}\right]  ^{q_{m}}\prod_{l=1}\left[  \left(  l-1\right)  !\frac
{\tilde{t}^{\prime}}{2M_{2}}\right]  ^{r_{l}}\nonumber\\
\cdot &  \sum_{i}^{q_{1}}\binom{q_{1}}{i}\left(  \frac{-s}{\tilde{t}}\right)
^{i}\sum_{j}^{r_{1}}\binom{r_{1}}{j}\left(  \frac{-s}{\tilde{t}^{\prime}%
}\right)  ^{j}\left(  -1\right)  ^{-i-j}B\left(  \frac{-t}{2}-1+i+j,\frac
{-u}{2}-1\right)
\end{align}
In the Regge limit, $t=$ fixed and $s\sim-u\rightarrow\infty,$ one gets%

\begin{align}
R_{tu}^{\left\vert p_{n},q_{m},r_{l}\right\rangle }\simeq &  \left(
-1\right)  ^{N}\prod_{n=1}\left[  \left(  n-1\right)  !\sqrt{-t}\right]
^{p_{n}}\prod_{m=1}\left[  \left(  m-1\right)  !\frac{\tilde{t}}{2M_{2}%
}\right]  ^{q_{m}}\prod_{l=1}\left[  \left(  l-1\right)  !\frac{\tilde
{t}^{\prime}}{2M_{2}}\right]  ^{r_{l}}\nonumber\\
\cdot &  F_{1}\left(  -\frac{t}{2}-1;-q_{1},-r_{1};\frac{s}{2};-\frac
{s}{\tilde{t}},-\frac{s}{\tilde{t}^{\prime}}\right)  B\left(  -\frac{t}%
{2}-1,-\frac{u}{2}-1+N\right)  .
\end{align}
Finally, one can derive the string BCJ relation in the Regge limit
\begin{equation}
\frac{R_{st}^{\left\vert p_{n},q_{m},r_{l}\right\rangle }}{R_{tu}^{\left\vert
p_{n},q_{m},r_{l}\right\rangle }}=\frac{\left(  -1\right)  ^{N}B\left(
-\frac{s}{2}-1,-\frac{t}{2}-1\right)  }{B\left(  -\frac{t}{2}-1,-\frac{u}%
{2}-1+N\right)  }=\frac{\sin\pi\left(  \frac{u}{2}+2-N\right)  }{\left(
-1\right)  ^{N}\sin\pi\left(  \frac{s}{2}+2\right)  }=\frac{\sin k_{2}\cdot
k_{4}}{\sin k_{1}\cdot k_{2}}. \label{RR2}%
\end{equation}
In constrast to the linear relations calculated in the hard scattering limit
in Eq.(\ref{HS}), it was shown \cite{LY2014,LY} that there existed infinite
recurrence relation among RSS amplitudes. For example, the recurrence relation
\cite{LY2014} ($(N;q_{1},r_{1})$ etc. refer to states in Eq.(\ref{GR}))%
\begin{equation}
\sqrt{-t}\left[  R_{st}^{(N;q_{1},r_{1})}+R_{st}^{(N;q_{1}-1,r_{1}+1)}\right]
-MR_{st}^{(N;q_{1}-1,r_{1})}=0 \label{APP}%
\end{equation}
for arbitrary mass levels $M^{2}=2(N-1)$ can be derived from recurrence
relations of the Appell functions. Eq.(\ref{RR2}) and Eq.(\ref{APP}) can be
combined to form one example of the \textit{extended recurrence relations in
the RSS limit}. The possible connection of field theory BCJ relations
\cite{BCJ} and Regge string recurrence relations \cite{LY} was first suggested
in \cite{LY}. These relations relate string scattering amplitudes of string
states with different spins and different channels in the RSS limit. Similar
to the HSS limit, it can be shown \cite{LY2014,LY} that the complete extended
recurrence relations in the RSS limit can be used to reduce the infinite
number of independent RSS amplitudes from $\infty$ down to $1$.

We have seen in this section that the HSS and the RSS amplitudes calculated
previously are consistent with the string BCJ relation in Eq.(\ref{BCJ3}). In
the next section, we will give an explicit proof the string BCJ relation for
all energies. In addition in section IV, we will calculate the nonrelativistic
limit of string BCJ relation to obtain the extended recurrence relation in the
nonrelativistic scattering limit.%

%TCIMACRO{\TeXButton{equation number}{\setcounter{equation}{0}
%\renewcommand{\theequation}{\arabic{section}.\arabic{equation}}}}%
%BeginExpansion
\setcounter{equation}{0}
\renewcommand{\theequation}{\arabic{section}.\arabic{equation}}%
%EndExpansion

\section{Explicit Proof of String BCJ}

In this section, we generalize the explicit calculations of high energy four
point string BCJ relations reviewed in the last section to all energy. The
proof of $n$-point string BCJ relations using monodromy was given in
\cite{stringBCJ} without calculating string amplitudes, Here we will
explicitly calculate string scattering amplitudes for four arbitrary string
states for both $s-t$ and $t-u$ channels and directly prove the four point
string BCJ relations.

There are at least two motivations to calculate the string BCJ relation
explicitly and give an alternative proof of the relations. Firstly, the proof
in \cite{stringBCJ} assumed\ negative real parts of $k_{i}\cdot k_{j}$ , and
puts some constraints on the kinematic regime for the validity of the string
BCJ relations. Our explicit proof here is on the contrary valid for all
kinematic regimes. Secondly, in section II the explicit calculations of
scattering amplitudes in the high energy string BCJ relation had led to the
extended relations both in the hard scattering limit and the Regge limit. As
we will see in section IV, the explicit calculation of the string BCJ relation
in the nonrelativistic scattering limit will also lead to new recurrence
relations among low energy string scattering amplitudes. Similarly, we will
see along the calculation of this section that the equality of the string BCJ
relations can be identified as the equalities of coefficients of two
multi-linear polynomials of\ ${k_{1}^{\mu}}$ and ${k_{3}^{\nu}}$ in the $s-t$
and $t-u$ channel amplitudes.

Instead of path integral approach, we will use the method of Wick contraction
to do the open string scattering amplitude calculation. As usual, we will be
fixing $SL(2,R)$ by choosing string worldsheet coordinates to be
$x_{1}=0,x_{3}=1,x_{4}=\infty$. We first give the answer of a simple example
($\alpha^{\prime}=\frac{1}{2};s-t$ channel)
\begin{align}
\mathcal{T}^{\mu\nu}  &  =\int%
%TCIMACRO{\tprod _{i=1}^{4}}%
%BeginExpansion
{\textstyle\prod_{i=1}^{4}}
%EndExpansion
dx_{i}<e^{ik_{1}X}\partial^{2}X^{(\mu}\partial X^{\nu)}e^{ik_{2}X}e^{ik_{3}%
X}e^{ik_{4}X}>\nonumber\\
&  =\frac{\Gamma(-\frac{s}{2}-1)\Gamma(-\frac{t}{2}-1)}{\Gamma(\frac{u}{2}%
+2)}\left[  \frac{t}{2}\left(  {\frac{t^{2}}{4}-1}\right)  {k_{1}^{\mu}%
k_{1}^{\nu}-}\left(  {\frac{s}{2}+1}\right)  {\frac{t}{2}}\left(  {\frac{t}%
{2}+1}\right)  {k_{1}^{\mu}k_{3}^{\nu}}\right. \nonumber\\
&  \left.  {+\frac{s}{2}}\left(  {\frac{s}{2}+1}\right)  \left(  {\frac{t}%
{2}+1}\right)  {k_{3}^{\mu}k_{1}^{\nu}-\frac{s}{2}}\left(  {\frac{s^{2}}{4}%
-1}\right)  {k_{3}^{\mu}k_{3}^{\nu}}\right]  .
\end{align}
The result is a multi-linear polynomial of\ ${k_{1}^{\mu}}$ and ${k_{3}^{\nu}%
}$ due to the choice of worldsheet coordinates above. To prove the equality of
$s-t$ and $t-u$ channel calculation, we can just show the equality
of\ coefficients of a typical term in each channel.

There are two key observations before we proceed to do the calculation.
Firstly, we can drop out the fourth vertex $V_{4}(x_{4})$ in the real
calculation due to the choice $x_{4}=\infty.$ Secondly, there are two types of
contributions in the contractions between two vertex operators. They are
contraction between $\partial^{a}X$ and $\partial^{a^{\prime}}X$, and
contraction between $\partial^{a}X$ and $e^{ikX}$.

we are going to calculate the most general four point function of string
vertex
\begin{align}
&  \left\langle V_{1}(x_{1})V_{2}(x_{2})V_{3}(x_{3})V_{4}(x_{4})\right\rangle
\nonumber\\
=  &  \left\langle \overset{a}{\left(  \partial X\right)  }\overset{b}{\left(
\partial X\right)  }\overset{c}{\left(  \partial X\right)  }%
\overset{d}{\left(  \partial X\right)  }e^{ik_{1}X}(x_{1})\overset{e}{\left(
\partial X\right)  }\overset{d^{\prime}}{\left(  \partial X\right)
}\overset{f}{\left(  \partial X\right)  }\overset{g}{\left(  \partial
X\right)  }e^{ik_{2}X}(x_{2})\overset{g^{\prime}}{\left(  \partial X\right)
}\overset{h}{\left(  \partial X\right)  }\overset{i}{\left(  \partial
X\right)  }\overset{b^{\prime}}{\left(  \partial X\right)  }e^{ik_{3}X}%
(x_{3})V_{4}(x_{4})\right\rangle .
\end{align}
We can write down the relavent three vertex operators as
\begin{subequations}
\label{vv}%
\begin{align}
V_{1}(x_{1})  &  =\overset{a}{\left(  \partial X\right)  }\overset{b}{\left(
\partial X\right)  }\overset{c}{\left(  \partial X\right)  }%
\overset{d}{\left(  \partial X\right)  }e^{ik_{1}X}(x_{1}),\\
V_{2}(x_{2})  &  =\overset{e}{\left(  \partial X\right)  }\overset{d^{\prime
}}{\left(  \partial X\right)  }\overset{f}{\left(  \partial X\right)
}\overset{g}{\left(  \partial X\right)  }e^{ik_{2}X}(x_{2}),\\
V_{3}(x_{3})  &  =\overset{g^{\prime}}{\left(  \partial X\right)
}\overset{h}{\left(  \partial X\right)  }\overset{i}{\left(  \partial
X\right)  }\overset{b^{\prime}}{\left(  \partial X\right)  }e^{ik_{3}X}%
(x_{3}),
\end{align}
where
\end{subequations}
\begin{subequations}
\label{vvv}%
\begin{align}
\overset{a}{\left(  \partial X\right)  }  &  =\prod_{a=1}^{A}\left(
i\varepsilon_{11a}^{(a)}\cdot\partial_{1}^{\alpha_{11a}}X_{1}\right)
,\overset{b}{\left(  \partial X\right)  }=\prod_{b=1}^{B}\left(
i\varepsilon_{12b}^{(b)}\cdot\partial_{1}^{\alpha_{12b}}X_{1}\right)
,\overset{c}{\left(  \partial X\right)  }=\prod_{c=1}^{C}\left(
i\varepsilon_{13c}^{(c)}\cdot\partial_{1}^{\alpha_{13c}}X_{1}\right)  ,\\
\overset{d}{\left(  \partial X\right)  }  &  =\prod_{d=1}^{D}\left(
i\varepsilon_{14d}^{(d)}\cdot\partial_{1}^{\alpha_{14d}}X_{1}\right)
,\overset{e}{\left(  \partial X\right)  }=\prod_{e=1}^{E}\left(
i\varepsilon_{21a}^{(e)}\cdot\partial_{2}^{\alpha_{21e}}X_{2}\right)
,\overset{d^{\prime\prime}}{\left(  \partial X\right)  }=\prod_{d^{\prime}%
=1}^{D}\left(  i\varepsilon_{22b}^{(d^{\prime})}\cdot\partial_{2}%
^{\alpha_{22d^{\prime}}}X_{2}\right)  ,\\
\overset{f}{\left(  \partial X\right)  }  &  =\prod_{f=1}^{F}\left(
i\varepsilon_{23f}^{(f)}\cdot\partial_{2}^{\alpha_{23f}}X_{2}\right)
,\overset{g}{\left(  \partial X\right)  }=\prod_{g=1}^{G}\left(
i\varepsilon_{24g}^{(g)}\cdot\partial_{2}^{\alpha_{24g}}X_{2}\right)
,\overset{g^{\prime}}{\left(  \partial X\right)  }=\prod_{g^{\prime}=1}%
^{G}\left(  i\varepsilon_{31e}^{(e^{\prime})}\cdot\partial_{3}^{\alpha
_{31g^{\prime}}}X_{3}\right)  ,\\
\overset{h}{\left(  \partial X\right)  }  &  =\prod_{h=1}^{H}\left(
i\varepsilon_{32h}^{(h)}\cdot\partial_{3}^{\alpha_{32h}}X_{3}\right)
,\overset{i}{\left(  \partial X\right)  }=\prod_{i=1}^{I}\left(
i\varepsilon_{33i}^{(i)}\cdot\partial_{3}^{\alpha_{33i}}X_{3}\right)
,\overset{b^{\prime}}{\left(  \partial X\right)  }=\prod_{b^{\prime}=1}%
^{B}\left(  i\varepsilon_{34b}^{(b^{\prime})}\cdot\partial_{3}^{\alpha_{34b}%
}X_{3}\right)  .
\end{align}
In Eq.(\ref{vvv}), $\varepsilon_{11a}^{(a)}$ and $\alpha_{11a}$ etc. are
polarizations and orders of worldsheet differential respectively. In
Eq.(\ref{vv}), we have four groups of $\partial^{a}X$ for each vertex
operator. Two of them will contract with $\partial^{b}X$ of the other
two\ vertex operators respectively, and the rest two will contract with
$e^{ikX}$ of the other two vertex operators respectively. For illustration, we
have used the pair dummy indexes $b,b^{\prime};d,d^{\prime}$ and $g,g^{\prime
}$ for contractions. The other six indexes $a,c,e,f,h$ and $i$ are prepared
for contractions with $e^{ikX}.$

The mass levels of the three vertex operators are%

\end{subequations}
\begin{subequations}
\begin{align}
N_{1}  &  =S_{A}+\text{\ }S_{B}+S_{C}+S_{D},\\
N_{2}  &  =S_{E}+S_{D}^{\prime}+S_{F}+S_{G},\\
N_{3}  &  =S_{G}^{\prime}+S_{H}+S_{I}+S_{B}^{\prime},
\end{align}
where we defined%
\end{subequations}
\begin{subequations}
\begin{align}
S_{A}  &  =\sum_{a=1}^{A}\alpha_{11a}\text{, \ \ }S_{B}=\sum_{b=1}^{B}%
\alpha_{12b}\text{, \ \ }S_{C}=\sum_{c=1}^{C}\alpha_{13c}\text{, \ \ }%
S_{D}=\sum_{d=1}^{D}\alpha_{14d},\\
S_{E}  &  =\sum_{e=1}^{E}\alpha_{21e}\text{, \ \ }S_{F}=\sum_{f=1}^{F}%
\alpha_{23f}\text{, \ \ }S_{G}=\sum_{g=1}^{G}\alpha_{24g}\text{, \ \ }%
S_{H}=\sum_{h=1}^{H}\alpha_{32h},\\
S_{I}  &  =\sum_{i=1}^{I}\alpha_{33i}\text{, \ \ }S_{B}^{\prime}%
=\sum_{b^{\prime}=1}^{B}\alpha_{34b^{\prime}}\text{, \ \ }S_{D}^{\prime}%
=\sum_{d^{\prime}=1}^{D}\alpha_{22d^{\prime}}\text{, \ \ }S_{G}^{\prime}%
=\sum_{g^{\prime}=1}^{G}\alpha_{31g^{\prime}}.
\end{align}
Then we have%
\end{subequations}
\begin{subequations}
\begin{align}
s  &  =-(k_{1}+k_{2})^{2},t=-(k_{2}+k_{3})^{2},u=-(k_{1}+k_{3})^{2},\\
k_{1}\cdot k_{2}  &  =-\frac{s}{2}-2+N_{1}+N_{2},\\
k_{2}\cdot k_{3}  &  =-\frac{t}{2}-2+N_{2}+N_{3},\\
k_{1}\cdot k_{3}  &  =-\frac{u}{2}-2+N_{1}+N_{3},\\
s+t+u  &  =2N^{\prime}-8\text{, \ \ with\ }N^{\prime}=N_{1}+N_{2}+N_{3}+N_{4}.
\end{align}
We are now ready to do the calculation. After putting the $SL(2,R)$ gauge
choice, we get%

\end{subequations}
\begin{align}
T=  &  \prod_{a=1}^{A}\left[  \varepsilon_{11a}^{(a)}\cdot k_{3}%
(-1)(\alpha_{11a}-1)!\right]  \prod_{b=1}^{B}\left[  \varepsilon_{12b}%
^{(b)}\cdot\varepsilon_{34b}^{(b)}(-1)^{\alpha_{34b}-1}(\alpha_{12b}%
+\alpha_{34b}-1)!\right] \nonumber\\
\cdot &  \prod_{c=1}^{C}\left[  \varepsilon_{13c}^{(c)}\cdot k_{2}%
(-1)(\alpha_{13c}-1)!\right]  \prod_{d=1}^{D}\left[  \varepsilon_{14d}%
^{(d)}\cdot\varepsilon_{22d}^{(d)}(-1)^{\alpha_{22d}-1}(\alpha_{14d}%
+\alpha_{22d}-1)!\right] \nonumber\\
\cdot &  \prod_{e=1}^{E}\left[  \varepsilon_{21e}^{(e)}\cdot k_{1}%
(-1)^{^{\alpha_{21e}}}(\alpha_{21e}-1)!\right]  \prod_{f=1}^{F}\left[
\varepsilon_{23f}^{(f)}\cdot k_{3}(-1)(\alpha_{23f}-1)!\right] \nonumber\\
\cdot &  \prod_{g=1}^{G}\left[  \varepsilon_{24g}^{(g)}\cdot\varepsilon
_{31g}^{(g)}(-1)^{\alpha_{31g}-1}(\alpha_{24g}+\alpha_{31g}-1)!\right]
\prod_{h=1}^{H}\left[  \varepsilon_{32h}^{(h)}\cdot k_{2}(-1)(\alpha
_{32h}-1)!\right] \nonumber\\
\cdot &  \prod_{i=1}^{I}\left[  \varepsilon_{33i}^{(i)}\cdot k_{1}%
(-1)^{\alpha_{33i}}(\alpha_{33i}-1)!\right] \nonumber\\
\cdot &  \int dx|-x|^{k_{1}\cdot k_{2}}|x-1|^{k_{2}\cdot k_{3}}x^{-\left(
S_{C}+S_{D}+S_{D}^{\prime}+S_{E}\right)  }\left(  1-x\right)  ^{-\left(
S_{F}+S_{G}+S_{G}^{\prime}+S_{H}\right)  }. \label{long}%
\end{align}
It is easy to see that only the last term of Eq.(\ref{long}) will be different
for the $s-t$ and $t-u$ channel calculations. For the $s-t$ channel, the last
term becomes%
\begin{align}
&  \int_{0}^{1}dx\text{ }x^{k_{1}\cdot k_{2}}\left(  1-x\right)  ^{k_{2}\cdot
k_{3}}x^{-\left(  S_{C}+S_{D}+S_{D}^{\prime}+S_{E}\right)  }\left(
1-x\right)  ^{-\left(  S_{F}+S_{G}+S_{G}^{\prime}+S_{H}\right)  }\nonumber\\
=  &  \int_{0}^{1}dx\times x^{-\frac{s}{2}-2+S_{A}+S_{B}+S_{F}+S_{G}%
}(1-x)^{-\frac{t}{2}-2+S_{E}+S_{D}+S_{I}+S_{B}^{\prime}}\nonumber\\
=  &  \frac{\Gamma\left(  \frac{-s}{2}-1+S_{A}+S_{B}+S_{F}+S_{G}\right)
\Gamma\left(  -\frac{t}{2}-1+S_{E}+S_{D}^{\prime}+S_{I}+S_{B}^{\prime}\right)
}{\Gamma\left(  \frac{u}{2}+2-N_{4}-S_{C}-S_{D}-S_{G}^{\prime}-S_{H}\right)
}.
\end{align}
For the $t-u$ channel, we have the last term%
\begin{equation}
\int_{1}^{\infty}dx(x)^{k_{1}\cdot k_{2}}(x-1)^{k_{2}\cdot k_{3}}x^{-\left(
S_{C}+S_{D}+S_{D}^{\prime}+S_{E}\right)  }(1-x)^{-\left(  S_{F}+S_{G}%
+S_{G}^{\prime}+S_{H}\right)  }.
\end{equation}
Define%

\begin{equation}
\text{ }K=-\left(  S_{F}+S_{G}+S_{G}^{\prime}+S_{H}\right)  ,
\end{equation}
and make the chane of variable $x=\frac{1}{1-y}$ in the integration, we end up with%

\begin{align}
&  (-1)^{K}\int_{0}^{1}dy(y)^{\frac{-t}{2}-2+S_{E}+S_{D}+S_{I}+S_{B}^{\prime}%
}(1-y)^{-\frac{u}{2}-2+N_{4}+S_{C}+S_{D}+S_{G}^{\prime}+S_{H}}\nonumber\\
=  &  (-1)^{K}\frac{\Gamma\left(  \frac{-t}{2}-1+S_{E}+S_{D}+S_{I}%
+S_{B}^{\prime}\right)  \Gamma\left(  -\frac{u}{2}-1+N_{4}+S_{C}+S_{D}%
+S_{G}^{\prime}+S_{H}\right)  }{\Gamma\left(  \frac{s}{2}+2-S_{A}-S_{B}%
-S_{F}-S_{G}\right)  }.
\end{align}
We are now ready to calculate the ratio%
\begin{align}
\frac{T_{st}}{T_{tu}}  &  =\frac{(-1)^{K}\Gamma\left(  \frac{-s}{2}%
-1+S_{A}+S_{B}+S_{F}+S_{G}\right)  \Gamma\left(  \frac{s}{2}+2-S_{A}%
-S_{B}-S_{F}-S_{G}\right)  }{\Gamma\left(  -\frac{u}{2}-1+N_{4}+S_{C}%
+S_{D}+S_{G}^{\prime}+S_{H}\right)  \Gamma\left(  \frac{u}{2}+2-N_{4}%
-S_{C}-S_{D}-S_{G}^{\prime}-S_{H}\right)  }\nonumber\\
&  =\left(  -1\right)  ^{-\left(  S_{F}+S_{G}+S_{G}^{\prime}+S_{H}\right)
}\cdot\frac{\sin\pi\left(  \frac{u}{2}+2-N_{4}-S_{C}-S_{D}-S_{G}^{\prime
}-S_{H}\right)  }{\sin\pi\left(  \frac{s}{2}+2-S_{A}-S_{B}-S_{F}-S_{G}\right)
}\nonumber\\
&  =\left(  -1\right)  ^{-N_{1}-N_{4}}\frac{\sin\pi\left(  \frac{u}%
{2}+2\right)  }{\sin\pi\left(  \frac{s}{2}+2\right)  }=\frac{\sin\pi\left(
k_{2}\cdot k_{4}\right)  }{\sin\pi\left(  k_{1}\cdot k_{2}\right)  },
\end{align}
where we have used the identity (\ref{math}). We thus have provd the four
point string BCJ relation by explicit calculation.%

%TCIMACRO{\TeXButton{equation number}{\setcounter{equation}{0}
%\renewcommand{\theequation}{\arabic{section}.\arabic{equation}}}}%
%BeginExpansion
\setcounter{equation}{0}
\renewcommand{\theequation}{\arabic{section}.\arabic{equation}}%
%EndExpansion

\section{Nonrelativistic String BCJ and Extended Recurrence Relations}

In this section, in constrast to the two high energy limits of string BCJ
relations discussed in section II, we discuss mass level dependent
nonrelativistic string BCJ relations. For simplicity, we will first calculate
both $s-t$ and $t-u$ channel NSS amplitudes of three tachyons and one leading
trojectory string state at arbitrary mass levels. We will then calculate NSS
amplitudes of three tachyons and one more general string state. We will see
that the mass and spin dependent nonrelativistic string BCJ relations can be
expressed in terms of Gauss hypergeometry functions. As an application, for
each fixed mass level $N$ we can then derive extended recurrence relations
among NSS amplitudes of string states with different spins and different channels.

\bigskip We choose $k_{2}$ to be momentum of the leading trojectory string
states and the rest are tachyons. In the CM frame%
\begin{subequations}
\begin{align}
k_{1}  &  =\left(  \sqrt{M_{1}^{2}+\vec{k_{1}}^{2}},-|\vec{k_{1}}|,0\right)
,\\
k_{2}  &  =\left(  \sqrt{M_{2}^{2}+\vec{k_{1}}^{2}},+|\vec{k_{1}}|,0\right)
,\\
k_{3}  &  =\left(  \sqrt{M_{3}^{2}+\vec{k_{3}}^{2}},-|\vec{k_{3}}|\cos
\phi,-|\vec{k_{3}}|\sin\phi\right)  ,\\
k_{4}  &  =\left(  \sqrt{M_{3}^{2}+\vec{k_{3}}^{2}},+|\vec{k_{3}}|\cos
\phi,+|\vec{k_{3}}|\sin\phi\right)
\end{align}
where $M_{1}=M_{3}=M_{4}=M_{tachyon}$ , $M_{2}=2(N-1)$ and $\phi$ is the
scattering angle on the scattering plane. Instead of the zero slope limt which
was used in the literature to get the field theory limit of the lowest mass
string state \cite{ymzero1,ymzero2}, \cite{Bzero1,Bzero2,Bzero3}, we will take
the nonrelativistic $|\vec{k_{1}}|<<M_{S}$ or large $M_{S}$ limit for the
massive string scattering amplitudes. In the nonrelativistic limit
($|\vec{k_{1}}|<<M_{S}$)%

\end{subequations}
\begin{subequations}
\begin{align}
k_{1}\simeq &  \left(  M_{1}+\frac{\vec{k_{1}}^{2}}{2M_{1}},-|\vec{k_{1}%
}|,0\right)  ,\\
k_{2}\simeq &  \left(  M_{2}+\frac{\vec{k_{1}}^{2}}{2M_{2}},+|\vec{k_{1}%
}|,0\right)  ,\\
k_{3}\simeq &  \left(  -\frac{M_{1}+M_{2}}{2}-\frac{1}{4}\frac{M_{1}+M_{2}%
}{M_{1}M_{2}}|\vec{k_{1}}|^{2},-\left[  \frac{\epsilon}{2}+\frac{(M_{1}%
+M_{2})}{4M_{1}M_{2}\epsilon}|\vec{k_{1}}|^{2}\right]  \cos\phi,\right.
\nonumber\\
&  \left.  -\left[  \frac{\epsilon}{2}+\frac{(M_{1}+M_{2})}{4M_{1}%
M_{2}\epsilon}|\vec{k_{1}}|^{2}\right]  \sin\phi\right)  ,\\
k_{4}\simeq &  \left(  -\frac{M_{1}+M_{2}}{2}-\frac{1}{4}\frac{M_{1}+M_{2}%
}{M_{1}M_{2}}|\vec{k_{1}}|^{2},+\left[  \frac{\epsilon}{2}+\frac{(M_{1}%
+M_{2})}{4M_{1}M_{2}\epsilon}|\vec{k_{1}}|^{2}\right]  \cos\phi,\right.
\nonumber\\
&  \left.  +\left[  \frac{\epsilon}{2}+\frac{(M_{1}+M_{2})}{4M_{1}%
M_{2}\epsilon}|\vec{k_{1}}|^{2}\right]  \sin\phi\right)  .
\end{align}
where%

\end{subequations}
\begin{equation}
\epsilon=\sqrt{(M_{1}+M_{2})^{2}-4M_{3}^{2}}.
\end{equation}
The three polarizations on the scattering plane are defined to be
\cite{ChanLee1,ChanLee2}%

\begin{subequations}
\begin{align}
e^{P}  &  =\frac{1}{M_{2}}\left(  \sqrt{M_{2}^{2}+\vec{k_{1}}^{2}},|\vec
{k_{1}}|,0\right)  ,\\
e^{L}  &  =\frac{1}{M_{2}}\left(  |\vec{k_{1}}|,\sqrt{M_{2}^{2}+\vec{k_{1}%
}^{2}},0\right)  ,\\
e^{T}  &  =(0,0,1),
\end{align}
which in the low energy limit reduce to%

\end{subequations}
\begin{subequations}
\begin{align}
e^{P}  &  \simeq\frac{1}{M_{2}}\left(  M_{2}+\frac{\vec{k_{1}}^{2}}{2M_{2}%
},|\vec{k_{1}}|,0\right)  ,\\
e^{L}  &  \simeq\frac{1}{M_{2}}\left(  |\vec{k_{1}}|,M_{2}+\frac{\vec{k_{1}%
}^{2}}{2M_{2}},0\right)  ,\\
e^{T}  &  \simeq\left(  0,0,1\right)  .
\end{align}
One can then calculate the\ following kinematics which will be used in the low
energy amplitude calculation%

\end{subequations}
\begin{subequations}
\begin{align}
k_{1}\cdot e^{L}  &  =k_{1}^{L}=\frac{-(M_{1}+M_{2})}{M_{2}}|\vec{k_{1}%
}|+O\left(  |\vec{k_{1}}|^{2}\right)  ,\\
k_{3}\cdot e^{L}  &  =k_{3}^{L}=\frac{-\epsilon}{2}\cos\phi+\frac{M_{1}+M_{2}%
}{2M_{2}}|\vec{k_{1}}|+O\left(  |\vec{k_{1}}|^{2}\right)  ,\\
k_{1}\cdot e^{T}  &  =k_{1}^{T}=0,\\
k_{3}\cdot e^{T}  &  =k_{3}^{T}=\frac{-\epsilon}{2}\sin\phi+O\left(
|\vec{k_{1}}|^{2}\right)  ,\\
k_{1}\cdot e^{P}  &  =k_{1}^{P}=-M_{1}+O\left(  |\vec{k_{1}}|^{2}\right)  ,\\
k_{3}\cdot e^{P}  &  =k_{3}^{P}=\frac{M_{1}+M_{2}}{2}-\frac{\epsilon}{2M_{2}%
}\cos\phi|\vec{k_{1}}|+O\left(  |\vec{k_{1}}|^{2}\right)  ,
\end{align}
amd the Mandelstam variables%
\end{subequations}
\begin{subequations}
\begin{align}
s  &  =\left(  M_{1}+M_{2}\right)  ^{2}+O\left(  |\vec{k_{1}}|^{2}\right)  ,\\
t  &  =-M_{1}M_{2}-2-\epsilon\cos\phi|\vec{k_{1}}|+O\left(  |\vec{k_{1}}%
|^{2}\right)  ,\\
u  &  =-M_{1}M_{2}-2+\epsilon\cos\phi|\vec{k_{1}}|+O\left(  |\vec{k_{1}}%
|^{2}\right)  .
\end{align}

\subsection{Leading Trojectory States}

We first calculate the nonrelativistic $s-t$ channel scattering amplitude of
three tachyons and one tensor string state%

\end{subequations}
\begin{equation}
V_{2}=(i\partial X^{T})^{p}(i\partial X^{L})^{r}(i\partial X^{P})^{q}%
e^{ik_{2}X}.
\end{equation}
where%
\begin{equation}
N=p+r+q.
\end{equation}
To the leading order in energy, the nonrelativistic amplitude can be
calculated to be,%

\begin{align}
A_{st}^{(p,r,q)}=  &  \int_{0}^{1}dx\left(  \frac{k_{1}^{T}}{x}-\frac
{k_{3}^{T}}{1-x}\right)  ^{p}\left(  \frac{k_{1}^{L}}{x}-\frac{k_{3}^{L}}%
{1-x}\right)  ^{r}\left(  \frac{k_{1}^{P}}{x}-\frac{k_{3}^{P}}{1-x}\right)
^{q}|x|^{k_{1}\cdot k_{2}}|x-1|^{k_{2}\cdot k_{3}}\nonumber\\
\simeq &  \int_{0}^{1}dx\left(  \frac{\frac{\epsilon}{2}\sin\phi}{1-x}\right)
^{p}\left(  \frac{\frac{\epsilon}{2}\cos\phi}{1-x}\right)  ^{r}\left(
-\frac{M_{1}}{x}-\frac{\frac{M_{1}+M_{2}}{2}}{1-x}\right)  ^{q}x^{k_{1}\cdot
k_{2}}(1-x)^{k_{2}\cdot k_{3}}\nonumber\\
=  &  \left(  \frac{\epsilon}{2}\sin\phi\right)  ^{p}\left(  \frac{\epsilon
}{2}\cos\phi\right)  ^{r}\left(  -\frac{M_{1}+M_{2}}{2}\right)  ^{q}%
\cdot\overset{q}{\underset{l=0}{\sum}}\binom{q}{l}\left(  \frac{2M_{1}}%
{M_{1}+M_{2}}\right)  ^{l}\nonumber\\
\cdot &  B\left(  1-M_{1}M_{2}-l,\frac{M_{1}M_{2}}{2}+l\right) \nonumber\\
=  &  \left(  \frac{\epsilon}{2}\sin\phi\right)  ^{p}\left(  \frac{\epsilon
}{2}\cos\phi\right)  ^{r}\left(  -\frac{M_{1}+M_{2}}{2}\right)  ^{q}B\left(
1-M_{1}M_{2},\frac{M_{1}M_{2}}{2}\right) \nonumber\\
\cdot &  \overset{q}{\underset{l=0}{\sum}}\left(  -1\right)  ^{l}\binom{q}%
{l}\left(  \frac{2M_{1}}{M_{1}+M_{2}}\right)  ^{l}\frac{\left(  \frac
{M_{1}M_{2}}{2}\right)  _{l}}{\left(  M_{1}M_{2}\right)  _{l}}.
\end{align}
Finally the summation above can be performed to get the Gauss hypergeometry
function $_{2}F_{1}$,%

\begin{align}
A_{st}^{(p,r,q)}  &  =\left(  \frac{\epsilon}{2}\sin\phi\right)  ^{p}\left(
\frac{\epsilon}{2}\cos\phi\right)  ^{r}\left(  -\frac{M_{1}+M_{2}}{2}\right)
^{q}B\left(  1-M_{1}M_{2},\frac{M_{1}M_{2}}{2}\right) \nonumber\\
&  \cdot\text{ }_{2}F_{1}\left(  \frac{M_{1}M_{2}}{2};-q;M_{1}M_{2}%
;\frac{2M_{1}}{M_{1}+M_{2}}\right)  .
\end{align}
Similarly, we calculate the corresponding nonrelativistic $t-u$ channel
amplitude as,%
\begin{align}
A_{tu}^{\left(  p,r,q\right)  }=  &  \int_{1}^{\infty}dx\left(  \frac
{k_{1}^{T}}{x}-\frac{k_{3}^{T}}{1-x}\right)  ^{p}\left(  \frac{k_{1}^{L}}%
{x}-\frac{k_{3}^{L}}{1-x}\right)  ^{r}\left(  \frac{k_{1}^{P}}{x}-\frac
{k_{3}^{P}}{1-x}\right)  ^{q}|x|^{k_{1}\cdot k_{2}}|x-1|^{k_{2}\cdot k_{3}%
}\nonumber\\
\simeq &  \left(  -1\right)  ^{N}\left(  \frac{\epsilon}{2}\sin\phi\right)
^{p}\left(  \frac{\epsilon}{2}\cos\phi\right)  ^{r}\left(  -\frac{M_{1}+M_{2}%
}{2}\right)  ^{q}B\left(  \frac{M_{1}M_{2}}{2},\frac{M_{1}M_{2}}{2}\right)
\nonumber\\
&  \cdot\text{ }_{2}F_{1}\left(  \frac{M_{1}M_{2}}{2};-q;M_{1}M_{2}%
;\frac{2M_{1}}{M_{1}+M_{2}}\right)  .
\end{align}
We are now ready to calculate the ratio of $s-t$ and $t-u$ channel amplitudes,%

\begin{align}
\frac{A_{st}^{(p,r,q)}}{A_{tu}^{(p,r,q)}}  &  =\left(  -1\right)  ^{N}%
\frac{B\left(  -M_{1}M_{2}+1,\frac{M_{1}M_{2}}{2}\right)  }{B\left(
\frac{M_{1}M_{2}}{2},\frac{M_{1}M_{2}}{2}\right)  }\nonumber\\
&  =(-1)^{N}\frac{\Gamma\left(  M_{1}M_{2}\right)  \Gamma\left(  -M_{1}%
M_{2}+1\right)  }{\Gamma\left(  \frac{M_{1}M_{2}}{2}\right)  \Gamma\left(
-\frac{M_{1}M_{2}}{2}+1\right)  }\simeq\frac{\sin\pi\left(  k_{2}\cdot
k_{4}\right)  }{\sin\pi\left(  k_{1}\cdot k_{2}\right)  }, \label{NBCJ}%
\end{align}
where, in the nonrelativistic limit, we have,%

\begin{subequations}
\begin{align}
k_{1}\cdot k_{2}  &  \simeq-M_{1}M_{2},\\
k_{2}\cdot k_{4}  &  \simeq\frac{\left(  M_{1}+M_{2}\right)  M_{2}}{2}.
\end{align}
So we have ended up with a consistent nonrelativistic\textit{ level }$M_{2}%
$\textit{ dependent string BCJ relations}. Similar relations for $t-u$ and
$s-u$ channel amplitudes can be calculated. We stress that the above relation
is the stringy generalization of the massless field theory BCJ relation to the
higher spin stringy particles. Moreover, as we will show now that, there exist
much more relations among these amplitudes.

There existed a recurrence relation of Gauss hypergeometry function%
\end{subequations}
\begin{equation}
_{2}F_{1}(a;b;c;z)=\frac{c-2b+2+(b-a-1)z}{(b-1)(z-1)}\text{ }_{2}%
F_{1}(a;b-1;c;z)+\frac{b-c-1}{(b-1)(z-1)}\text{ }_{2}F_{1}(a;b-2;c;z),
\label{rec}%
\end{equation}
which can be used to derive the recurrence relation,%
\begin{align}
\left(  -\frac{M_{1}+M_{2}}{2}\right)  A_{st}^{(p,r,q)}=  &  \frac
{M_{2}\left(  M_{1}M_{2}+2q+2\right)  }{\left(  q+1\right)  \left(
M_{2}-M_{1}\right)  }\left(  \frac{\epsilon}{2}\sin\phi\right)  ^{p-p^{\prime
}}\left(  \frac{\epsilon}{2}\cos\phi\right)  ^{p^{\prime}-p+1}A_{st}^{\left(
p^{\prime},p+r-p^{\prime}-1,q+1\right)  }\nonumber\\
+  &  \frac{2\left(  M_{1}M_{2}+q+1\right)  }{\left(  q+1\right)  \left(
M_{2}-M_{1}\right)  }\left(  \frac{\epsilon}{2}\sin\phi\right)  ^{p-p^{\prime
\prime}}\left(  \frac{\epsilon}{2}\cos\phi\right)  ^{p^{\prime\prime}%
-p+2}A_{st}^{\left(  p^{\prime\prime},p+r-p^{\prime\prime}-2,q+2\right)  }
\label{main}%
\end{align}
where $p^{\prime}$ and $p^{\prime\prime}$ are the polarization parameters of
the second and third Amplitudes on the rhs of Eq.(\ref{main}). For example,
for a fixed mass level $N=4,$ one can derive many recurrence relations for
either $s-t$ channel or $t-u$ channel amplitudes with $q=0,1,2.$ For say
$q=2,$ $(p,r)=(2,0),(1,1),(0,2).$ We have $p^{\prime}=0,1$ and $p^{\prime
\prime}=0.$ We can thus derive, for example for $(p,r)=(2,0)$ and $p^{\prime
}=1$, the recurrence relation among NSS amplitudes $A_{st}^{(2,0,2)}%
A_{st}^{(1,0,3)}A_{st}^{(0,0,4)}$ as following
\begin{equation}
\left(  -\frac{M_{1}+M_{2}}{2}\right)  A_{st}^{(2,0,2)}=\frac{M_{2}\left(
M_{1}M_{2}+6\right)  }{3\left(  M_{2}-M_{1}\right)  }\left(  \frac{\epsilon
}{2}\sin\phi\right)  A_{st}^{(1,0,3)}+\frac{2\left(  M_{1}M_{2}+4\right)
}{3\left(  M_{2}-M_{1}\right)  }\left(  \frac{\epsilon}{2}\sin\phi\right)
^{2}A_{st}^{(0,0,4)}.
\end{equation}
Exactly the same relation can be obtained for $t-u$ channel amplitudes since
the $_{2}F_{1}(a;b;c;z)$ dependence in the $s-t$ and $t-u$ channel amplitudes
calculated above are the same. Moreover, we can for example replace
$A_{st}^{(2,0,2)}$ amplitude above by the corresponding $t-u$ channel
amplitude $A_{tu}^{(2,0,2)}$ through Eq.(\ref{NBCJ}) and obtain%
\begin{align}
\frac{\left(  -1\right)  ^{N}}{2\cos\frac{\pi M_{1}M_{2}}{2}}\left(
-\frac{M_{1}+M_{2}}{2}\right)  A_{tu}^{(2,0,2)}  &  =\frac{M_{2}\left(
M_{1}M_{2}+6\right)  }{3\left(  M_{2}-M_{1}\right)  }\left(  \frac{\epsilon
}{2}\sin\phi\right)  A_{st}^{(1,0,3)}\nonumber\\
&  +\frac{2\left(  M_{1}M_{2}+4\right)  }{3\left(  M_{2}-M_{1}\right)
}\left(  \frac{\epsilon}{2}\sin\phi\right)  ^{2}A_{st}^{(0,0,4)}, \label{BCJJ}%
\end{align}
which relates higher spin NSS amplitudes in both $s-t$ and $t-u$ channels.
Eq.(\ref{BCJJ}) is one example of the \textit{extended recurrence relations in
the NSS limit}. For each fixed mass level $M_{2}$, the relation in
Eq.(\ref{BCJJ}) relates amplitudes of different spin polarizations and
different channels of the \textit{same} propogating higher spin particle in
the string spectrum. In the next subsection, we will consider a more general
extended recurrence relation which relates NSS amplitudes of
\textit{different} higher spin particles for each fixed mass level\ $M_{2}$ in
the string spectrum.

\subsection{More general string states}

Recently the structure of the most general NSS string amplitudes which can be
expressed in terms of Gauss hypergeometry functions were pointed out
\cite{LLY}. Here, as an illustration, we will calculate one example of
extended recurrence relation which relates NSS amplitudes of different higher
spin particles for each fixed mass level\ $M_{2}$. In particular, the $s-t$
channel of NSS amplitudes of three tachyons and one higher spin massive string
state at mass level $N=3p_{1}+q_{1}+3$ correspond to the following three
higher spin string states%

\begin{align}
&  A_{1}\symbol{126}\left(  i\partial^{3}X^{T}\right)  ^{p_{1}}\left(
i\partial X^{P}\right)  ^{1}\left(  i\partial X^{L}\right)  ^{q_{1}+2},\\
&  A_{2}\symbol{126}\left(  i\partial^{2}X^{T}\right)  ^{p_{1}}\left(
i\partial X^{P}\right)  ^{2}\left(  i\partial X^{L}\right)  ^{p_{1}+q_{1}%
+1},\\
&  A_{3}\symbol{126}\left(  i\partial X^{T}\right)  ^{p_{1}}\left(  i\partial
X^{P}\right)  ^{3}\left(  i\partial X^{L}\right)  ^{2p_{1}+q_{1}}%
\end{align}
can be calculated to be%
\begin{align}
A_{1}  &  =\left[  2!\frac{\epsilon}{2}\sin\phi\right]  ^{p_{1}}\left[
-\left(  1-1\right)  !\frac{M_{1}+M_{2}}{2}\right]  ^{1}\left[  0!\frac
{\epsilon}{2}\cos\phi\right]  ^{q_{1}+2}\nonumber\\
&  \times B\left(  \frac{M_{1}M_{2}}{2},1-M_{1}M_{2}\right)  \text{ }_{2}%
F_{1}\left(  \frac{M_{1}M_{2}}{2},-1,M_{1}M_{2},\frac{-2M_{1}}{M_{1}+M_{2}%
}\right)  ,\\
A_{2}  &  =\left[  1!\frac{\epsilon}{2}\sin\phi\right]  ^{p_{1}}\left[
-\left(  2-1\right)  !\frac{M_{1}+M_{2}}{2}\right]  ^{2}\left[  0!\frac
{\epsilon}{2}\cos\phi\right]  ^{p_{1}+q_{1}+1}\nonumber\\
&  \times B\left(  \frac{M_{1}M_{2}}{2},1-M_{1}M_{2}\right)  \text{ }_{2}%
F_{1}\left(  \frac{M_{1}M_{2}}{2},-2,M_{1}M_{2},\frac{-2M_{1}}{M_{1}+M_{2}%
}\right)  ,\\
A_{3}  &  =\left[  0!\frac{\epsilon}{2}\sin\phi\right]  ^{p_{1}}\left[
-\left(  3-1\right)  !\frac{M_{1}+M_{2}}{2}\right]  ^{3}\left[  0!\frac
{\epsilon}{2}\cos\phi\right]  ^{2p_{1}+q_{1}}\nonumber\\
&  \times B\left(  \frac{M_{1}M_{2}}{2},1-M_{1}M_{2}\right)  \text{ }_{2}%
F_{1}\left(  \frac{M_{1}M_{2}}{2},-3,M_{1}M_{2},\frac{-2M_{1}}{M_{1}+M_{2}%
}\right)  .
\end{align}
To apply the recurrence relation in Eq.(\ref{rec}) for Gauss hypergeometry
functions, we choose%
\begin{equation}
a=\frac{M_{1}M_{2}}{2},b=-1,c=M_{1}M_{2},z=\frac{-2M_{1}}{M_{1}+M_{2}}.
\end{equation}
One can then calculate the extended recurrence relation%
\begin{align}
&  16\left(  \frac{2M_{1}}{M_{1}+M_{2}}+1\right)  \left(  -\frac{M_{1}+M_{2}%
}{2}\right)  ^{2}\left(  \frac{\epsilon}{2}\cos\phi\right)  ^{2p_{1}}%
A_{1}\nonumber\\
&  =8\cdot2^{P_{1}}\left(  \frac{M_{1}M_{2}}{2}+2\right)  \left(  \frac
{2M_{1}}{M_{1}+M_{2}}+2\right)  \left(  -\frac{M_{1}+M_{2}}{2}\right)  \left(
\frac{\epsilon}{2}\cos\phi\right)  ^{p_{1}+1}A_{2}\nonumber\\
&  -2^{P_{1}}\left(  M_{1}M_{2}+2\right)  \left(  \frac{\epsilon}{2}\cos
\phi\right)  ^{2}A_{3}%
\end{align}
where $p_{1}$ is an arbitrary integer. More extendened recurrence relations
can be similarly derived.

The existence of these low energy stringy symmetries comes as a surprise from
Gross's high energy symmetries \cite{GM,Gross,GrossManes} point of view.
Finally, in constrast to the Regge string spacetime symmetry which\ was shown
to be related to $SL(5,C)$ of the Appell function $F_{1}$ \cite{LY2014}, here
we found that the low energy stringy symmetry is related to $SL(4,C)$
\cite{sl4c} of the Gauss hypergeometry functions $_{2}F_{1}.$

\section{Conclusion}

In this paper, we review historically two independent approaches of the four
point string BCJ relation. One originates from field theory BCJ relations
\cite{BCJ}, and the other from calculation of string scattering amplitudes in
the HSS limit \cite{Closed}. By combining string BCJ relations with infinite
linear relations of HSS amplitudes \cite{ChanLee1,ChanLee2,PRL, CHLTY,susy},
one obtains extended linear relations which relate HSS amplitudes of string
states with different spins and different channels. Moreover, these extended
linear relations can be used to reduce the number of independent HSS
amplitudes from $\infty$ down to $1$\cite{ChanLee1,ChanLee2,PRL, CHLTY,susy}.
Similar calculation can be performed in the RSS limit \cite{KLY}, and one
obtains extended recurrence relations in the RSS limit. These extended Regge
recurrence relations again can be used to reduce the number of independent RSS
amplitudes from $\infty$ down to $1$ \cite{LY,LY2014}.

We then give an explicit proof of four point string BCJ relations for all
energy. We found that the equality of the string BCJ relations can be
identified as the equalities of coefficients of two multi-linear polynomials
of\ ${k_{1}^{\mu}}$ and ${k_{3}^{\nu}}$ in the $s-t$ and $t-u$ channel
amplitudes. This calculation, which puts no constraints on the kinematic
regimes in constrast to the previous one \cite{BCJ}, provides an alternative
proof of the one based on monodromy of integration \cite{BCJ} in string
amplitude calculation.

Finally, we calculate both $s-t$ and $t-u$ channel NSS amplitudes of three
tachyons and one higher spin string state including the leading trojectory
string states at arbitrary mass levels. We discover that the mass and spin
dependent nonrelativistic string BCJ relations can be expressed in terms of
Gauss hypergeometry functions. As an application, we calculate examples of
extended recurrence relations of low energy NSS amplitudes. For each fixed
mass level $N,$ these extended recurrence relations relate low energy NSS
amplitudes of string states with different spins and different channels.

We believe that many string theory origins of properties of field theory
amplitudes remain to be understood, and many more stringy generalizations of
properties of field theory amplitudes remain to be uncovered in the near future.

\section{Acknowledgments}

We would like to thank Chung-I Tan for discussions. J.C. thanks the useful
correspondence of authors of reference \cite{stringBCJ}. He also thanks
Chih-Hau Fu for bringing his attention to reference \cite{stringBCJ}. This
work is supported in part by the Ministry of Science and Technology and S.T.
Yau center of NCTU, Taiwan.

\end{document}